# Remote Chip-Scale Quantum Sensing of Magnetic Fields

Kfir Levi, Avital Giat, Liran Golan, Eliran Talker and Liron Stern
*Institute of Applied Physics, The Hebrew University of Jerusalem, Jerusalem, 91904, Israel*

**Abstract:** Quantum sensing is an ever-evolving research field describing the use of a quantum phenomenon to perform measurement of a physical quantity. Amongst different types of quantum sensors, atomic vapor-based quantum effects are extensively used to measure quantities such as time, velocity, acceleration, and electric and magnetic fields. Here, we propose and demonstrate remote quantum sensing using a chip-scale atomic vapor cell. Specifically, we remotely interrogate mm-scale micromachined vapor cells, and measure the ambient Earth's magnetic field at a standoff distance of ~10 meters and a sensitivity of ~1 pT/√Hz. Simultaneously we are able measure the distance between micro-cell and the interrogating system by means of time-of-flight measurements, thus correlating between position and magnetic field. Consequently, we provide a novel toolset to measure and map arbitrary, remote, and hard to access magnetic field in unshielded environments with high sensitivity and spatial resolution, paving the way to a variety of novel applications in diverse fields such as medicine, communication, defense, space-exploration, and quantum technologies.

**Introduction.** Metrology, the science of measurement, strives to push the limits of human ability to measure quantities such as time, temperature, electromagnetic fields, and mass. This ongoing global effort provides us with a reference "ruler" which is crucial both for understanding the fundamental nature of our universe, as well as enabling precision measurements which drive new technologies. Quantum sensors[1] play a crucial role in measuring various physical quantities that are the backbone of today's technological breakthroughs. By exploiting atomic quantum effects, these sensors can detect an extensive range of parameters, including time/frequency[2–5], electric fields[6,7], magnetic fields[8,9], voltage[10,11], rotation[12], and temperature[13–16]. Quantum sensors offers significant advantages, such as high sensitivity, high precision, traceability and the ability to conduct non-invasive measurements.

Recent years have shown significant advances in the field of chip-scale quantum sensors, in particular those based on micromachined atomic vapor-cells[17–19]. Allowing above room-temperature operation, such miniaturized devices allow unprecedented levels of performance when considering their size, weight, power consumption (SWaP), and cost[20]. Consequently, such devices dramatically extend the application-space of quantum sensors. Indeed, such devices have already allowed the demonstration of novel types of atomic clocks and optical frequency references, such as the chip-scale atomic clock (CSAC)[21], the photonically-integrated micro-comb-based clock[4] and the micro-comb frequency reference[22]. This progress has also led to the exciting development of microfabricated-based chip-scale magnetometry, enabling sub pT-level sensitivity in micro-cells that are only a few millimeters in size[23] opening the grounds to numerous applications in fields such as geology, medicine, and environmental monitoring[24]. Indeed, chip-scale atomic magnetometers have been utilized in applications in magnetoencephalography[25], including arrays of fiber-based actuated micromachined vapor cells allowing spatial mapping of magnetic-fields[26], as well as the utilization of such arrays in magnetic sensor-helmets allowing measurements of neural activity[27]. Several experimental manifestations of atomic-based quantum sensors are all-optical, thus suggesting the possibility to enable remote atomic-based quantum sensing. Indeed, such concept has been demonstrated in an inspiring experiment where a natural occurring atomic reservoir in earth mesosphere (90 km above sea-level) has been remotely interrogated to allow the mapping of the magnetic field in that location[28]. Such demonstration has been supplemented by lab-based experiments in table-top vapor studies[29].

Here, we propose and experimentally demonstrate remote atomic-based quantum sensing using micromachined vapor cells. We experimentally demonstrate such concept using a $^{87}$Rb mm-scale anodically bonded vapor cell, and by preforming all-optical unshielded remote magnetometry, which allows us to remotely map magnetic field at a standoff distance of 10 m and sensitivity floor of 1 pT/√Hz. Such mode of operation, offers novel and exciting opportunities, including applications where proximity is not feasible or when the object being measured is in a harsh or hazardous environment. Moreover, such concept may also provide a non-invasive method to measure magnetic fields, which can be important in certain applications where the object being measured is sensitive or fragile. The ability to mass-fabricate such cells at the wafer-scale, suggests the possibility of dispersing a myriad of cells, resembling dust, to allow remote spatial mapping of magnetic fields. Although our current demonstration is focused on magnetometry, such conceptual arrangement may be extended to other types of quantum sensors such as Rydberg electric field sensing. It is expected that such remote quantum sensing experimental apparatus may find numerous applications in several important fields. These may include defense applications (ordnance detection, perimeter monitoring), environmental monitoring, geological surveying, and medical diagnostics. Another potential envisioned application, is space exploration. Here, one may foresee the remote measurement of magnetic fields of celestial bodies, such as planets, moons, and asteroids which can have a significant impact on space missions and communication systems.

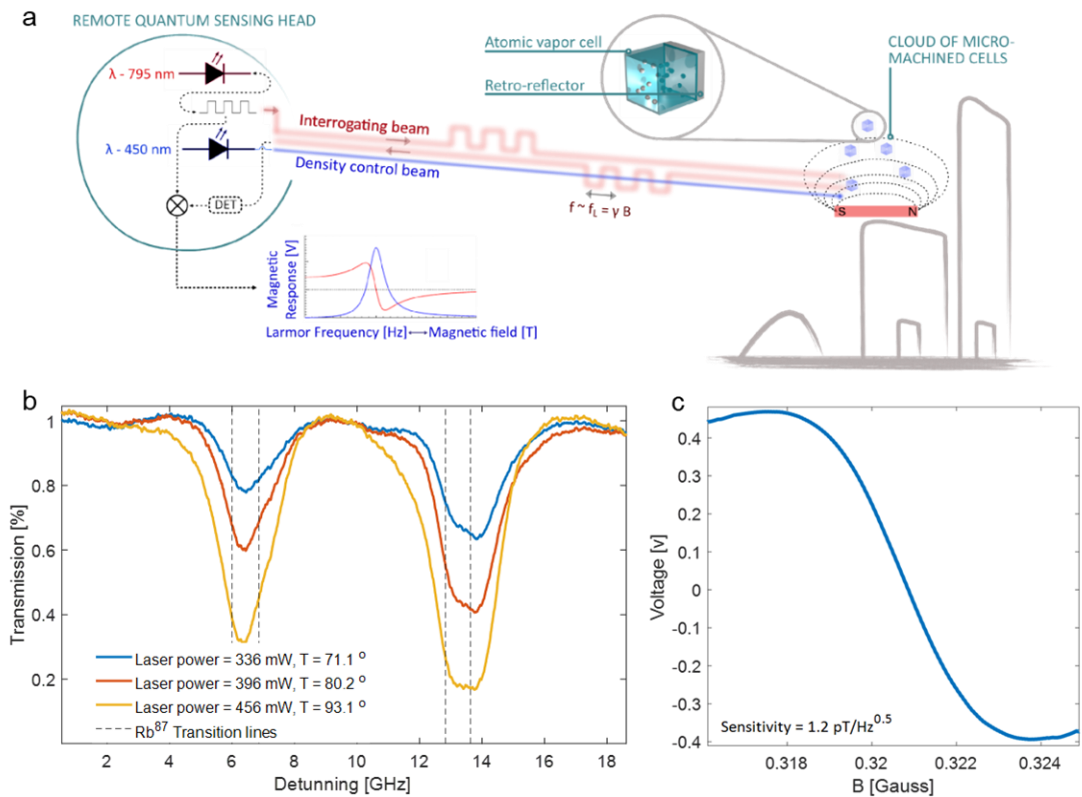

Figure 1 | Concept of remote micromachined based quantum sensing (a) Artistic deception of remote all optical magnetic sensing concept (b) Remote spectroscopy of D1 optical lines of $^{87}$Rb obtained by irradiating a micromachined vapor cell with a density control laser and an interrogating retroreflected laser (c) Remote magnetic resonance Bell-Bloom spectroscopic signal corresponding to a noise floor sensitivity of ~ 1 pT/hz$^{1/2}$ at Fourier frequencies of ~400 Hz.

**Concept of remote micromachined based quantum sensing.** Figure 1 presents a conceptual overview of our apparatus and results. First, we describe conceptually remote quantum magnetic sensing using a cloud of micromachined vapor cells. Two main components exist in this system. First, we illustrate a sensing head (coined, "remote quantum sensing head") containing a density control laser, modulated interrogating laser and detection opto-electronics. Subsequently, in a remote location (urban or ex-terrestrial location) we describe a cloud of micromachined vapor cells which are interrogated by the above-mentioned sensing head. A laser impinges on a specific cell, whilst modulated at the Larmor frequency ($f_L = \gamma B$, $\gamma$ being the gyromagnetic constant and B being the magnetic field) associated with the local magnetic field in the vicinity of the micro-cell. In order to increase and control the density of the atomic cell above its room-temperature value, an additional laser is used which effectively induces material absorption, and in turn heats the cell and increases the atomic density. The retroreflected modulated light is collected at the same sensing head located remotely from the cloud of atomic cells. Through synchronous lock-in detection, one can deduce the magnetic field at the vicinity of the micromachined cell in a remote and all-optical fashion. Two basic results are presented in this overview figure. First, we demonstrate our ability to control the atomic density using absorption of light at the atomic cell structure. Specifically, In Fig. 1b we illustrate the spectroscopic dependence of the $^{87}$Rb D1 lines at the power of a heating laser. Here, one clearly observes a strong dependence of the atomic optical density which corresponds to atomic densities varying from $1.5 \cdot 10^{10}$ cm$^{-3}$ (at room temperature and zero optical heating power) to $4 \cdot 10^{12}$ cm$^{-3}$ (corresponding to ~93 °C and ~460 mW optical power). With this ability in hand, we then turn to obtain a remote magnetic resonance dispersion signal which is presented in Fig. 1c. Here, we perform a synchronous Bell-

Bloom detection scheme. Thus, in means of modulation drive the system at the Larmor frequency corresponding to the local magnetic field in the vicinity of the micromachined cell. The presented data has been averaged to display the slow frequency response. As we will show explicitly in the following figures, the corresponding inferred minimal sensitivity of such measurements is at the $pT \cdot Hz^{-1/2}$ level for Fourier frequencies larger than of ~ 300 Hz.

A crucial part of the ability to provide efficient, cost-effective, and high-resolution remote quantum sensing are micromachined vapor cells. Therefore, we now turn to describe the micromachined cells used in this study. Our cells are fabricated by means of anodically bonding layers of Pyrex to processed silicon layers, illustrated in Fig. 2b. $^{87}$RB atoms are deposited in the cell together with a few tens of Torr of $N_2$ buffer gas allowing prolonged ground state life-times as well as providing the means to quench fluorescence. In our current realization the cells have mm-scale cross sections (3 X 3 X 5 mm³). Such cells can be further miniaturized, allowing thousands of cells to be produced at the wafer-level. It would be noted, however, that hand in hand with reduction in cell dimensions the sensitivity limit might be degraded. This is a consequence of the reduction in the number of participating atoms in as well as the probable decrease of lifetime. Yet, it is still expected that in the 100³ µm³ regime of cell volume, pT levels of signal sensitivity floor may be feasible. Moreover, wafer-level production technology may provide significant opportunities, by enabling the integration of sophisticated photonic elements within such micromachined vapor cells. These could allow, amongst many, optical confinement, multi-pass beam propagation and integrated retroreflection which will all dramatically enhance the capability of micro-cells.

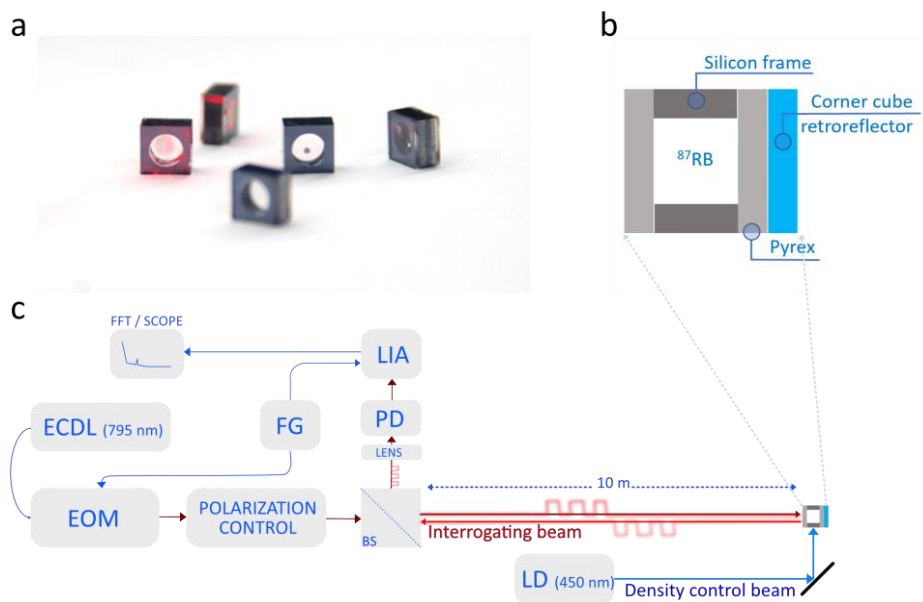

Figure 2 | Micromachined vapor-cells and optical arrangement (a) Micromachined mm-scale atomic vapor cells (b) Micromachined mm cell remote-sensing heterostructure (c) Remote all optical magnetic sensing schematic optical and electric arrangement; ECDL (External Cavity Diode Laser), EOM (Electro Optical Modulator) FG (Function Generator), LD (Laser Diode), LIA (Lock-In Amplifier).

**Remote Bell-Bloom magnetometry with mm-scale cells.** We now turn to describe the systems components in detail. The experimental setup is shown in Fig 2c. Here, an external cavity diode laser tuned to D1 $^{87}$Rb transition line is modulated by means of an electro optical modulator (EOM). Subsequently, we circularly polarize our interrogating beam. In this configuration, during the interaction with the rubidium atoms, the circularly polarized light aligns the magnetic moment of a gas of alkali $^{87}$Rb atoms through optical pumping[30]. In presence of an ambient DC magnetic field $B_0$ the atoms magnetic moment precesses around the magnetic field direction at the Larmor frequency $w_L = \gamma B_0$ where γ is the gyromagnetic moment. The modulator is used in order to achieve coherent spin precession[31] (when tuned to $w_L$) A mm-scale micromachined-cell (Pyrex-Si-Pyrex, anodic bonding Fig 2a, and 2b) filled with $^{87}$Rb atoms mixed with buffer gas is placed in a ~10 m distance from the interrogation unit. Since we want to measure an arbitrary magnetic field, the cell is in a completely unshielded environment. Owning to the cell's relatively small optical path, it is necessary to increase the atomic density. Density control is achieved using a visible diode laser (DL, 450 nm), placed a few meters from the cell, which irradiates and is absorbed in the cell's silicon frame. In principle it is possible to stabilize the atomic density, for instance, by feedbacking the atomic absorption. Yet, we have found that the time constants associated with the power fluctuations and thermal time constants allow one to "freely" irradiate the atoms with the heating laser. From the same reason, we didn't focus here on fully optimizing the distance between the visible diode heating laser and the micromachined cell, as the only parameter governing this distance is beam diffraction, i.e. it's Rayleigh length. This is in vast contrast with the interrogating beam (ECDL, ~795 nm) which propagates through free space, reflected after single pass through the cell and back, and finally propagates through free space to the detection unit. In this process signal integrity (amplitude, phase and polarization - noise) is crucial for the quality of our remote sensing. In our current manifestation of the system we use a corner-cube (CC) as the reflector, which allows relative broad acceptance angles with respect to specular reflection from a mirror. In order to detect the modulated retroreflected signal, we focus the reflected light into an amplified photodetector, which in turn is demodulated by means of a lock-in amplifier (LIA). The use of a lens allows us to collect more signal from the beam, even when we are strictly above the Rayleigh length. This configuration enables us to detect the signal amplitude and phase for a given modulation frequency. Operationally, we scan the modulation frequency, applied to the EOM and to the LIA and obtain the magnetic spectroscopic information (i.e., the magnetic field in vicinity of the micromachined cell) which features coincide with the Larmor frequency as a consequence of coherent spin precession generated in the atomic medium.

**Remote chip-scale magnetometry in earth's magnetic field.** We first explore the magnetic spectroscopic features of our remotely interrogated micromachine cell. To so, as mentioned above, we place our micromachined vapor cell at a standoff distance of ~10 m, and use ~ 500 µW and ~ 365 mW of interrogating and density-control laser powers. The latter corresponds to a density of $1.5 \cdot 10^{12}$ cm$^{-3}$ and temperature of approximately 75 °C. Both lasers have mm-scale diameters, which corresponds to Rayleigh lengths of 10.9 m (ECDL) and ~100 m (LD). Our interrogating laser of 795 nm is amplitude modulated with a square-wave at frequencies about the mean magnetic-field, which corresponds to a Larmor frequency of approximately 225 kHz (~320 mG). Demodulation and low-pass filtering at kHz rate provides us with typical out of phase dispersive-resonant line-shape showcased in Fig 3a. As mentioned above, we operate our magnetometry in an unshielded laboratory environment prone to a plethora of magnetic frequency components. These are clearly evident in the dispersive signal, whereas the strongest 50 Hz components can be easily recognized (as seen in the inset of Figure 3a) and superimposed on the dispersive signal. Following, we set the modulation frequency at the vicinity of the zero-crossing of the magnetic resonances and record the power spectral density (PSD) of the demodulated signal, by means of a fast Fourier transform (FFT) analyzer. Typical measurements are presented in Figure 3b. Indeed, multiple Fourier components are evident, corresponding to harmonics of 50 Hz as well as 60 Hz. These are all easily traced to the magnetic contents of different electronic equipment in the laboratory. Interestingly, we have exploited such magnetic frequency components to aid us in optimization of the sensitivity of our remote magnetic sensing.

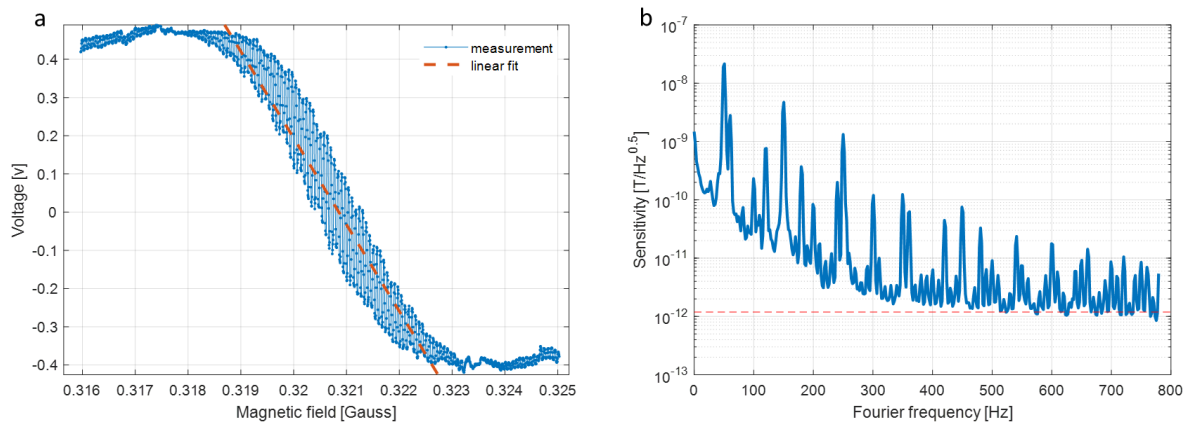

Figure 3 | **Remote magnetic resonance spectroscopy and sensitivity limit** (a) Remote magnetic resonance obtained by demodulating a 10 m retroreflected beam in the Bell-Bloom configuration; (b) The corresponding measured magnetic sensitivity noise floor spectral density demonstrating a noise floor of ~ 1 pT/√Hz at Fourier frequencies of ~300 Hz and above.

Magnetic dispersive signals such as presented in Fig. 3a are utilized to obtain the magnitude of response of our remote magnetometer in terms of volt deviations per unit magnetic field. In turn, such slope is used to calibrate the magnitude the measured PSDs, and consequently the systems ultimate sensitivity. Thus, in summary, we calculated the sensitivity as follows $Sensitivity = (dv/dB)^{-1}_{B=B_0} \cdot PSD_{w_m=w_L}(f > 300Hz)$. While $B_0 = w_L/\gamma$ is the environment DC magnetic field, $w_L$ is the Larmor frequency and $w_m$ is the modulation frequency. We find the noise floor of such measurements to be at the few pT/Hz$^{1/2}$ regime. A typical PSD is shown in Fig 3b. Here, the noise floor at ~ 300 Hz is $1.2 \frac{pT}{\sqrt{Hz}}$ with a demodulation bandwidth set to roughly coincide with the atomic bandwidth of ~1 kHz. The sensitivity of our system matches similar systems that measure the Earth's magnetic field [32]. Most likely the sensitivity is limited by the combination of the intrinsic coherence of our cell owning to, buffer gas composition as well at the limitations of measuring in ambient conditions such as magnetic fields gradients. Additionally, in order to verify these results, we applied (by means of Helmholtz coils) a given alternating magnetic field perpendicular to the propagation direction of light. This magnetic frequency components appeared in the Fourier PSD domain and allowed us to verify that our magnetic field calibration. We indeed witness a consistency of the signal appearing in the PSD with the calibrated noise floor given reasonable uncertainties in the calibrating field and recalling that we implemented here a scalar magnetometer, i.e., the alternating magnetic field magnitude measured is a fraction of the magnitude of the Larmor frequency.

**Exploring the sensitivity dependence on interrogating and heating lasers.** Important parameters for remote magnetic sensing are the dependence of the sensitivity on the incident powers of the interrogating and heating lasers. Beyond their operational significance, these can infer our ability to further extend the standoff distance for remote quantum chip-scale sensing. First, we measure the PSDs (i.e. the spectral sensitivity) as function of interrogating lasers power. Such typical dependence is presented in Figure 4. Clearly, the sensitivity is monotonically dependent on the power levels of the integrating laser. For instance, for a power of 17 µW (red curve, Fig 4) we have a sensitivity floor of about 0.2 nT Hz$^{-1/2}$, whereas for 565 µW we reach our current noise floor of ~1 pT Hz$^{-1/2}$. Careful examination of the power dependence sensitivity shows three different power laws. For low powers we witness a quadratic dependence, that evolves into a linear dependence and finally to a fractional (i.e, ~ P$^{-0.3}$) dependence. Such behavior can be explained by the combination of power broadening and the onset of the shot-noise limited regime.

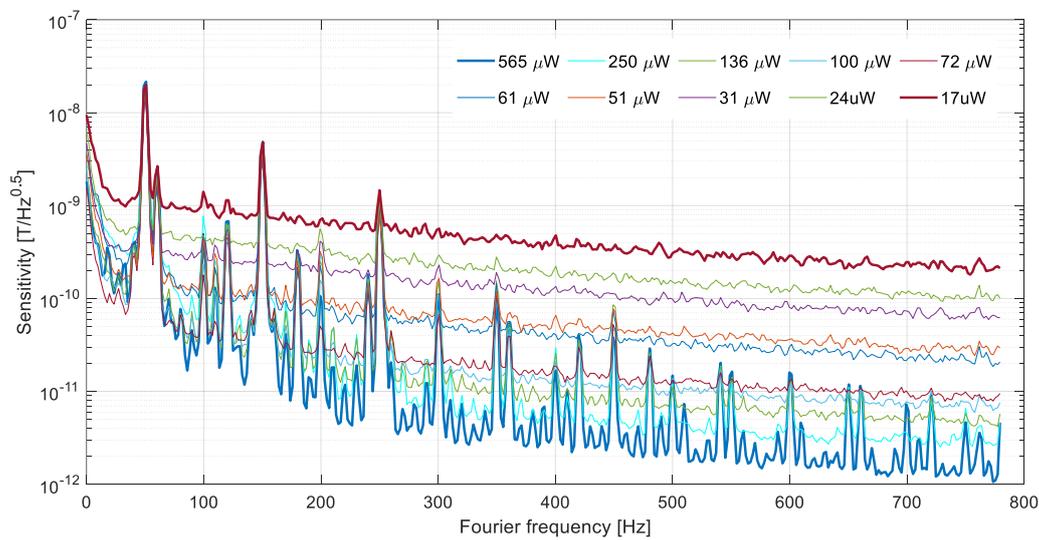

Figure 4 | **Sensitivity dependence on interrogating laser power** Sensitivity spectral densities for different interrogating laser powers varying from 17 µW to 0.565 mW.

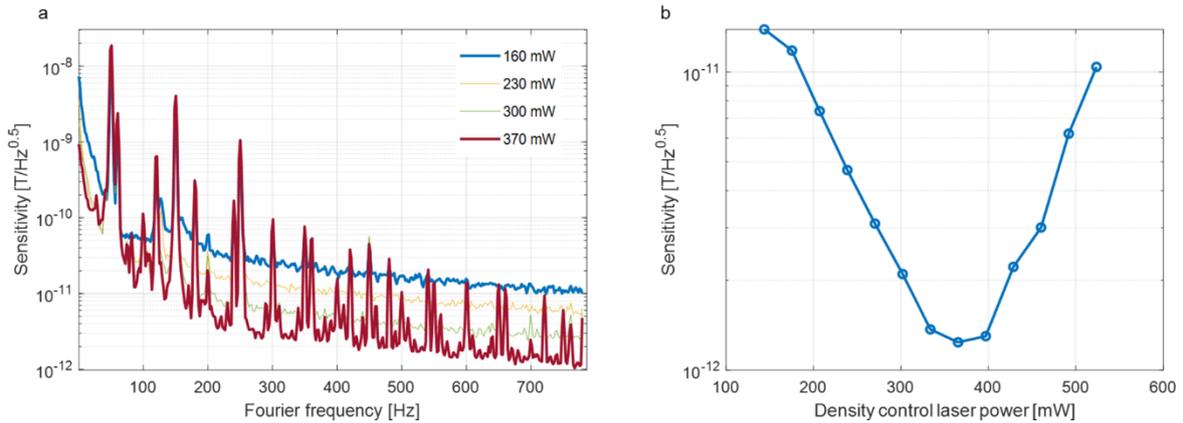

Figure 5 | Sensitivity dependence on heating laser power (a) Sensitivity spectral densities for different heating laser powers varying from 140 mW to 370 mW. (b) Sensitivity noise floor as function of density control laser power

Following, we explore the sensitivity of our system as function of the temperature of the cell, controlled by the remote blue laser diode. In figure 5a we plot a few representative spectral densities, as we increase the power of the heating lasers, and consequently the density of the atoms in the micromachined cell. The noise floor is plotted for various powers in Figure 5b. Clearly, an optimum sensitivity can be obtained at heating powers around 360 mW. Such behavior can be explained as follows: For low LD current, the signal directly follows the density of the atoms, and consequently the sensitivity improves. Yet, in some stage the increased absorption prevents signal to reach the detector and lowers the sensitivity. The cell temperature for this measurement is between 40 °C to 90 °C and the optimal cell temperature is about 75 °C which correspond to LD power of 365 mW. As evident in Figure 5b for a deviation of ±20 °C from the optimal temperature the sensitivity increases/decreases at most by an order of magnitude. Such relative broad operation-space has a profound effect on the ability to remotely interrogate the atomic medium and obtain sensing information.

**Correlating distance and magnetometry data.** Here, we present another mode of operation where we measure the distance between the atomic vapor cell and the detection unit. Such mode of operation may allow us to correlate between the quantum sensing information and the distance. Here, exploiting the electro-optic modulator already existing in our setup, we pulse-modulate the interrogating laser to allow a time-of-flight ranging measurement. A result of such arrangement is presented in Figure 6, where we measure the time of arrival of such pulse. We exploit a faint spurious reflection (Fig. 6, small pulse, at 0 ns) from our beam-splitter to calibrate the time of flight between the retro-reflection from the corner cube micromachine vapor cell (Fig. 6, larger pulse, with arrival time of ~62 ns). Thus, we determine the one-way distance to be 9.79 m. By employing a scanning laser system which tracks a reference emitting angle, accompanied with synchronized switching between magnetic field measurement operation mode to time-of-flight measurement operation mode one can remotely obtain spatial magnetic field maps.

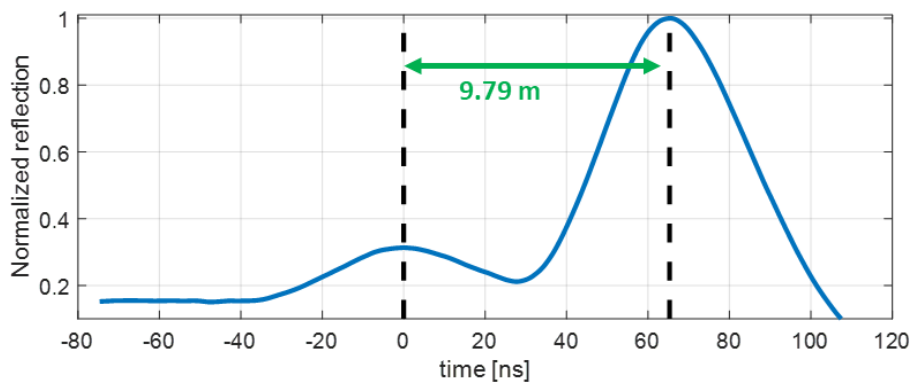

Figure 6 | Time of flight distance measurement Pulsed interrogating laser retroreflected from beam-splitter (0 ns) and micromachine vapor-cell corner-cube structure indicating a one-way path of 9.79 m.

**Discussion and summary:** In this study, have realized a remote atomic magnetic quantum sensor at standoff distance of 10 meters. It is natural to attempt to evaluate the performance of the system at greater distances. We anticipate that two primary factors, namely low-frequency noise (1/f noise), and beam divergence, may limit the performance of our system as the measurement distance increases. Specifically owning to turbulence and vibrations, as the measurement distance increases, 1/f noise is expected to rise, which can impede our system's sensitivity at low frequencies. Resolving this may necessitate an active noise-cancelation approach, as implemented for instance in time/frequency distribution links[5]. Yet, one can assess the impact of beam divergence, and resultant change in optical power needed to maintain the intensity at the cell position, as well as the ability to maintain signal integrity at the detector. Our current system employs an interrogating laser beam with a diameter of 3.33 mm, and a corresponding Rayleigh length of ~ 11 m. With these parameters, we achieve an intensity of 65 W/cm$^2$. For a beam diameter of 10.5 mm (and a corresponding Rayleigh length of ~109 m), and thus an expected sensitivity of ~ 40 pT/(√Hz) for an average power of 0.5 mW. In order to achieve a sensitivity ~ 1 pT/(√Hz) for measurements at greater distances, for instance ~100 m, would necessitate an intensity of at least 20 W/cm$^2$ corresponding to 2 mW of power. Additionally, one would need to compensate for the larger beam diameter in the collecting unit. Correspondingly, the density control laser has a diameter of ~9 mm and a Rayleigh length of ~100 m which in principle allows measurement for ~100 m distance, yet optimization of cell size, composition and beam divergence may reduce this power.

Here, we have demonstrated the feasibility of remote magnetic field measurement using a single mm-scale micromachined atomic vapor cell. In future studies, one may increase the number of cells as well as decrease their size and remotely map the magnetic field in the environment by measuring the field at the vicinity of each cell. To facilitate efficient scanning and measurement across multiple cells, one can construct "smart" vapor cells tailored for efficient remote heating and collection. For instance, by incorporating an absorptive glass to the cells front and a micromachined array of corner-cube retroflectors. Implementation of meta-surfaces to allow efficient retroreflection is also possible, following a recent study by the Faraon group[33]. This will allow to combine the density control and interrogating lasers into a single beam and simplifying the measurement process. Finally, recent proposals by Zhang et al.[34], of using non-linear light conversion to facilitate mirrorless lasing may allow achieving remote sensing without the need for a physical photonic retroreflector.

To summarize, our study showcases a demonstration of chip-scale all-optical remote magnetic sensing, utilizing micromachined vapor cells in an unshielded environment. Operationally, we have reached a sensitivity noise floor of ~1 pT/√Hz, mm-scale resolution and a standoff distance of ~10 meters. Consequently, our study introduces a new set of tools for remotely mapping magnetic fields in unshielded environments with high sensitivity and spatial resolution. Remarkably, this technique can be expanded to allow mapping of a myriad of atomic-based quantum sensors by mass-fabrication and subsequent dispersion of a cloud of micro-cells in space. Such mode of operation may enable remote spatial mapping of the sensing quantity leading to a diverse range of applications such as space exploration, ordnance detection, and geophysical surveying with sensitive environments without any risk of contamination, which could be especially valuable for medical and biological research. Such novel toolset may expand the horizons of traditional quantum sensors and pave the way to diverse application in medicine, communication and quantum technologies.

## Methods

### Cell fabrication

Micromachined vapor cells are fabricated by etching through-holes in silicon wafers. Subsequently, this silicon frame is anodically bonded to a borosilicate wafer, whereinafter $^{87}$Rb vapor is dispensed into the voids and encapsulated via an additional anodic bonding step in a buffer gas ($N_2$) environment.

### Optical Arrangement & Detection

An external cavity laser (TOPTICA DL-PRO) at 795 nm is modulated using fiber-Mach-Zehnder amplitude modulator (IX-BLUE, NIR-MX800-LN) biased at the linear response regime. The light is collimated using a ~ 3 mm diameter collimator an propagates through free-space and "folding" mirrors at a total distance of ~10 m. A micromachined vapor-cell placed at the very close vicinity of a corner cube retroreflector (Thorlabs PS977) reflects the signal which back-propagates and is focused to an amplified-Si detector. Simultaneously, we launch the density control laser (a longitudinal multimode laser-diode at 450 nm; TL L450P1600MM) to heat the silicon frame of the micromachined vapor cell. The signal is synchronously demodulated using a lock-in amplifier (SRS SR865A) referenced to the modulated frequency. We analyze the output signal using either an oscilloscope or (In particular for noise floor measurements) using a FFT spectrum analyzer (SRS SR760).

## Data Availability
The data and code that support the findings of this study are available from the corresponding author on reasonable request.

## Ethics Declarations
Competing interests
The authors declare no competing interests.

## Acknowledgments
The authors would like to thank Dr. John Kitching and Dr. Douglass Bopp for fruitful discussions. K.L is grateful for support from the Gutsman fellowship. We acknowledge the Israel Council for Higher Education for their generous support through various Quantum Initiative programs.

## References:

1. Degen, C. L., Reinhard, F. & Cappellaro, P. Quantum sensing. *Rev Mod Phys* **89**, (2017).
2. Liew, L. A. *et al.* Microfabricated alkali atom vapor cells. *Appl Phys Lett* **84**, 2694–2696 (2004).
3. McGrew, W. F. *et al.* Atomic clock performance enabling geodesy below the centimetre level. *Nature* (2018) doi:10.1038/s41586-018-0738-2.
4. Newman, Z. L. *et al.* Architecture for the photonic integration of an optical atomic clock. *Optica* **6**, 680 (2019).
5. Predehl, K. *et al.* A 920-kilometer optical fiber link for frequency metrology at the 19th decimal place. *Science (1979)* (2012) doi:10.1126/science.1218442.
6. Sedlacek, J. A. *et al.* Microwave electrometry with Rydberg atoms in a vapour cell using bright atomic resonances. *Nat Phys* (2012) doi:10.1038/nphys2423.
7. Simons, M. T., Gordon, J. A. & Holloway, C. L. Fiber-coupled vapor cell for a portable Rydberg atom-based radio frequency electric field sensor. *Appl Opt* (2018) doi:10.1364/ao.57.006456.
8. Budker, D. & Romalis, M. Optical magnetometry. *Nat Phys* **3**, 227–234 (2007).
9. Kominis, I. K., Kornack, T. W., Allred, J. C. & Romalis, M. v. A subfemtotesla multichannel atomic magnetometer. *Nature* **422**, 596–599 (2003).
10. Hamilton, C. A., Burroughs, C. J. & Benz, S. P. Josephson voltage standard-a review. *IEEE Transactions on Appiled Superconductivity* **7**, 3756–3761 (1997).
11. Teale, C., Sherman, J. & Kitching, J. Degenerate two-photon Rydberg atom voltage reference. *AVS Quantum Science* **4**, 024403 (2022).
12. Gustavson, T. L., Bouyer, P. & Kasevich, M. A. Precision rotation measurements with an atom interferometer gyroscope. *Phys Rev Lett* (1997) doi:10.1103/PhysRevLett.78.2046.
13. Kominis, I. K., Kornack, T. W., Allred, J. C. & Romalis, M. v. A subfemtotesla multichannel atomic magnetometer. *Nature* **422**, 596–599 (2003).
14. Kitching, J., Knappe, S. & Donley, E. A. Atomic sensors - A review. *IEEE Sensors Journal* vol. 11 1749–1758 Preprint at https://doi.org/10.1109/JSEN.2011.2157679 (2011).
15. Sedlacek, J., Schwettmann, A., Kübler, H. & Shaffer, J. P. Atom Based Vector Microwave Electrometry Using Rubidium Rydberg Atoms in a Vapor Cell. (2013) doi:10.1103/PhysRevLett.111.063001.
16. Truong, G.-W., Anstie, J. D., May, E. F., Stace, T. M. & Luiten, A. N. Accurate lineshape spectroscopy and the Boltzmann constant. *Nat Commun* **6**, 8345 (2015).
17. Kitching, J. Chip-scale atomic devices. *Appl Phys Rev* **5**, 31302 (2018).
18. Dyer, S. *et al.* Micro-machined deep silicon atomic vapor cells. *J Appl Phys* **132**, (2022).
19. McGilligan, J. P. *et al.* Micro-fabricated components for cold atom sensors. *Review of Scientific Instruments* vol. 93 Preprint at https://doi.org/10.1063/5.0101628 (2022).
20. Kitching, J. Chip-scale atomic devices. *Applied Physics Reviews* vol. 5 Preprint at https://doi.org/10.1063/1.5026238 (2018).
21. Knappe, S. *et al.* A microfabricated atomic clock. *Appl Phys Lett* **85**, 1460–1462 (2004).
22. Stern, L. *et al.* Direct Kerr frequency comb atomic spectroscopy and stabilization. *Sci Adv* **6**, eaax6230 (2020).
23. Shah, V., Knappe, S., Schwindt, P. D. D. & Kitching, J. Subpicotesla atomic magnetometry with a microfabricated vapour cell. *Nat Photonics* **1**, 649–652 (2007).
24. Budker, D., Kimball, D. F., Rochester, S. M., Yashchuk, V. v. & Zolotorev, M. Sensitive magnetometry based on nonlinear magneto-optical rotation. *Phys Rev A (Coll Park)* **62**, 043403 (2000).
25. Sander, T. H. *et al.* Magnetoencephalography with a chip-scale atomic magnetometer. *Biomed Opt Express* (2012) doi:10.1364/boe.3.000981.
26. Alem, O. *et al.* Magnetic field imaging with microfabricated optically-pumped magnetometers. *Opt Express* **25**, 7849 (2017).

27. Seymour, R. A. *et al.* Using OPMs to measure neural activity in standing, mobile participants. *Neuroimage* **244**, (2021).
28. Pedreros Bustos, F. *et al.* Remote sensing of geomagnetic fields and atomic collisions in the mesosphere. *Nat Commun* **9**, (2018).
29. Patton, B. *et al.* A remotely interrogated all-optical 87Rb magnetometer. *Appl Phys Lett* **101**, (2012).
30. Happer, W. *Optical pumping*. (1971).
31. William E. Bell & Arnold L. Bloom. *Optically Driven Spin Precession. Progr. Theoret. Phys. (Kyoto)* vol. 6 (1961).
32. Lenci, L., Barreiro, S., Valente, P., Failache, H. & Lezama, A. A magnetometer suitable for measurement of the Earth's field based on transient atomic response. *Journal of Physics B: Atomic, Molecular and Optical Physics* **45**, (2012).
33. Arbabi, A., Arbabi, E., Horie, Y., Kamali, S. M. & Faraon, A. Planar metasurface retroreflector. *Nat Photonics* (2017) doi:10.1038/nphoton.2017.96.
34. Zhang, R. *et al.* Stand-Off Magnetometry with Directional Emission from Sodium Vapors. *Phys Rev Lett* **127**, (2021).